\documentclass [12pt,a4paper]{article}

\usepackage{graphicx}
\usepackage{amsfonts}    \usepackage{amssymb}
\usepackage{amsmath}
\newcommand{\D}[2]{\frac{\partial #2}{\partial #1}}
\newcommand{\DD}[2]{\frac{\partial^2 #2}{\partial #1^2}}

\newcommand{\eps}{\varepsilon}

\begin{document}

\title{ Numerical Model For Vibration Damping Resulting From the
First Order Phase Transformations}

\author{L. X.  Wang $^1$
  \thanks{Corresponding Address:Tel.:+45 6550 1686, Fax:+45 6550 1660,
  E-mail:wanglinxiang$@$mci.sdu.dk}
 \quad and \quad Roderick
V.N. Melnik $^2$  \\
 $^1$  MCI, Faculty of Science and Engineering, \\
 University of Southern Denmark,\\
 Sonderborg, DK-6400, Denmark \\
 $^2$ Mathematical Modelling and Computational Sciences, \\
  Wilfrid Laurier University, \\
 75 University Avenue West, \\
 Waterloo, ON, Canada, N2L 3C5 \\
 }

\date{}

\maketitle


\begin{abstract}
 A numerical model is constructed for modelling macroscale damping effects induced
 by the first order martensite phase transformations in a shape memory alloy rod.
 The model is constructed on the basis of the modified Landau-Ginzburg
 theory that couples nonlinear mechanical and thermal fields.
The free energy function for the model is constructed as a double well function at
low temperature, such that the external energy can be absorbed during  the
phase  transformation and converted into thermal form.
The Chebyshev spectral methods are employed together with  backward
differentiation for the numerical
analysis of the problem. Computational experiments  performed for different
 vibration energies demonstrate the importance of taking into account  damping effects induced
by phase transformations.


\end{abstract}


\section{Introduction}

Shape Memory Alloy  (SMA) materials are able to  directly transduce thermal
energy  into mechanical and vice versa.  Their  unique properties
make them very attractive in many engineering applications, including mechanical and control
engineering, biomedicine, communication, robotics and so on  \cite{Birman1997}.
Among these applications,  dampers made from SMA for passive and semi-active
vibration damping, are quoted perhaps most frequently \cite{Song2006}. These
devices exhibit fairly complicated nonlinear (hysteretic) behaviour induced by martensite
phase transformations. An appropriate mathematical model involving phase transformations
and thermo-mechanical coupling is essential for a better understanding of the dynamic
 behavior of SMA dampers.

The damping effects of  SMA cannot be understood without incorporating into the
model the {\em dynamics} of the first order martensite transformations induced
by mechanical loadings, and hysteresis as a consequence.
SMAs have more than one martensite variants at low temperature that correspond
to the same elastic potential energy of the material stored via deformation.
When the martensite phase transformation is induced mechanically, part of the external
 energy will be consumed to transform the material from one variant to another,
 without increasing the total elastic energy stored in the material, but the deformation
will be oriented in a way favored by the external loadings.
The work done by the external loadings during the phase transformation process
cannot be restored reversely when the loading is removed, since it is converted into
thermal form, and will be dissipated eventually via, for example,
thermal dissipation.  This dissipation of mechanical energy due to phase transformation
can be measured by the hysteretic behaviour of the material under
mechanical loadings \cite{Juan2003,Humbeeck2003}.
Because of the nonlinear nature  of phase transformations
and the nonlinear coupling between the mechanical and thermal fields,  the dynamic
behavior of SMA dampers becomes very complex.

Some investigations have been  already carried out to understand the damping
 effects of SMA at either micro-scale or macro-scale
 \cite{Humbeeck2003,Piedboeuf1998,Chen2000}.
 Among many efforts,  it was shown in Ref.   \cite{Humbeeck2003,Masuda2002}
 ( and references  therein) that the damping effects of SMA are influenced by
the vibration frequency, heating/cooling  rate (temperature change rate), and vibration amplitude .
This influence is discussed and analyzed in terms of strain formation, interface moving,
internal friction, and other factors,  which are all considered either at
micro-scale or mesoscale.

For most current engineering applications of SMA dampers, damping effects due to hysteresis
is more pronounced at macroscale, and one needs to model the dynamical behaviour
and damping effects of SMAs at macroscale, which demands that the model should have the
 capability to capture all the contributions to vibration damping, particularly that due to
 hysteresis induced by cyclic mechanical loadings \cite{Juan2003,Masuda2002}.
Since vibration damping can be induced by hysteretic behaviour of the SMAs,
interface moving, internal friction, and  thermo-mechanical coupling, the model
for damping devices made from SMA  has to be constructed on the basis of dynamics
of the materials, which  involves the phase transformation and thermo-mechanical coupling.
Progressively, the impact induced phase transformation in SMA wires were
investigated in \cite{Chen2000,Masuda2002} where  the constitutive laws
were approximated linearly, and  the thermo-mechanical  coupling
was neglected.  Such models have obvious limitations.

In what follows, we attempt to overcome  such limitations by
better capturing the thermo-mechanical coupling, hysteresis, and
nonlinear nature of phase transformations with the Ginzburg-Landau theory,
originally discussed in the context of SMAs in \cite{Falk1980,Matus2004} and later
applied to model the dynamical behavior of SMA rods under dynamical loadings (e.g.,
\cite{Bubner1996,Bubner2000,Kloucek2004} and references therein). In particular, in this
paper we employ a mathematical model based on the modified
Ginzburg - Landau theory for the SMA damper, the dynamics of the SMA damper are described by
 Navier's equation with a non-convex constitutive relation.  The damper is connected with
a mass block with a given initial velocity.  The movement of the mass block is then
simulated as a single degree of freedom system, subject to the damping force
from the SMA rod.


\section{Mathematical Modelling}

\begin{figure}   \begin{center}
\includegraphics[scale=0.5]{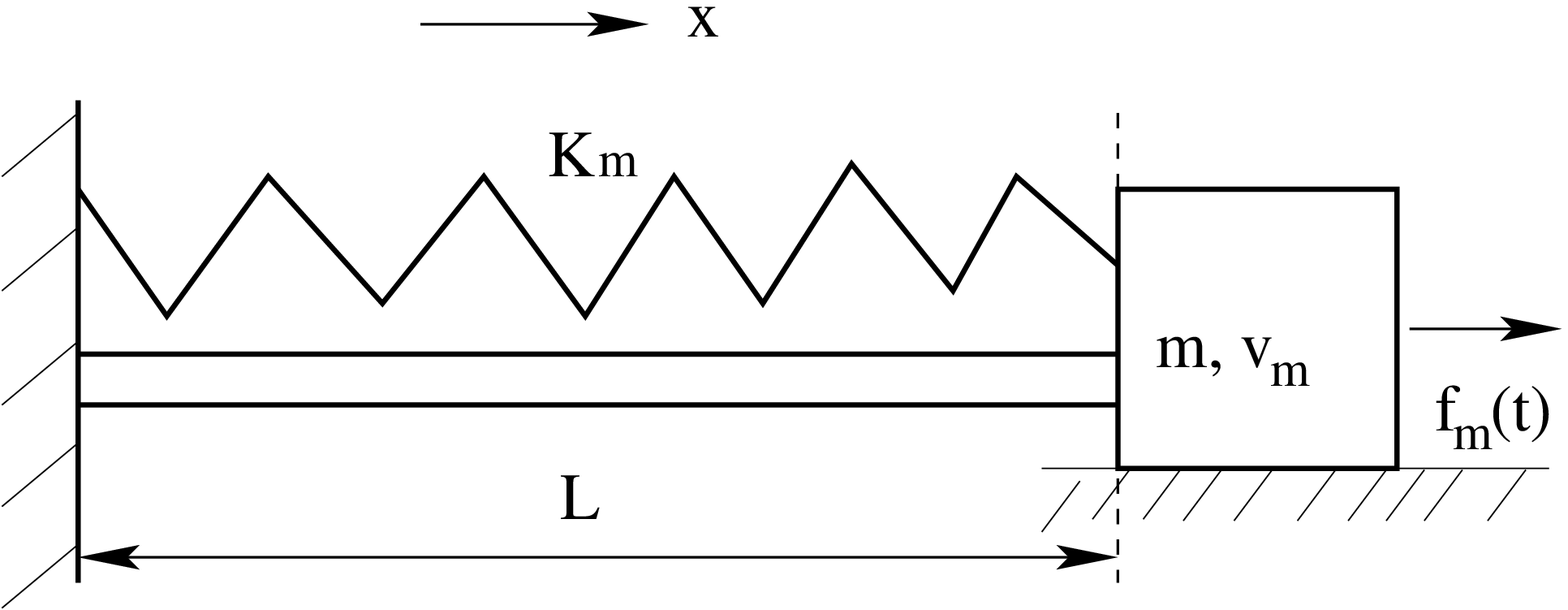}
\caption{Vibration damping of a mass block connected to
 a shape memory alloy rod.
\label{DampingSystem}     }
\end{center}    \end{figure}

The physical problem investigated here is sketched in Figure~(\ref{DampingSystem}).
There is a mass block connected to a SMA  rod. The SMA rod has a cross section
area of  $\beta$ and length $L$, and is initially at rest.  The block has an initial velocity
$v_m$. The purpose of the current investigation is to show the damping effect of the
SMA rod on  the vibration of the mass block.

In the current damping system, the mass block is a lumped sub-system,
while the SMA rod is a distributed sub-system.  Here, we construct the model
describing their
 movements separately and then couple them together. For the mass block,
the model can be easily formulated using the momentum conservation law as:
     \begin{equation}
       \label{MassEq}
	m \ddot x_m   + \mu_m \dot x_m + k_m (x_m - L ) = f_m,
	\end{equation}
\noindent where  $x_m$ is the position of the block, $\mu_m$  the friction coefficient,
 $k_m$  the spring stiffness,  $m$  the mass of the block, and $f_m$ is the external
 force on the mass block which is caused by the SMA rod discussed below.

The mathematical model for the dynamics of the SMA rod can be constructed on the
basis of the modified Ginzburg - Landau theory, as done in
(\cite{Falk1980,Matus2004,Bubner1996,Kloucek2004,Melnik2002} and references therein). For
the convenience of discussion about damping and energy absorbing here, the model is
constructed using the extended Hamilton's principle, but using the same non-convex potential
energy.

In order to formulate the governing equation for  the mechanical field,  the Lagrangean
 $\mathcal L$ is introduced for material particles as follows:
    \begin{eqnarray} \label{Lagrange}
     \mathcal L =  \frac{\rho}{2} (\dot u) ^2 - \mathcal F,
    \end{eqnarray}
where $\rho$ is the density of the material and $\mathcal F$ is the potential energy function
of the material.  The differntiating feature of the Ginzburg - Landau theory is that the potential
energy function is constructed as a non-convex function of the chosen \emph{order parameters}
and temperature $\theta$. It is a sum of local energy density $(\mathcal F_l)$ and non-local
energy density ($\mathcal F_g$). For the current one-dimensional  problem,
 the strain $\eps(x,t)=\D{x}{u}$  is  chosen as the order parameter,
 and the local free energy function
 can  be constructed as the Landau free energy function $\mathcal F_l(\theta, \eps)$
 \cite{Falk1980,Matus2004,Bubner1996,Bubner2000,Wang2004}:
    \begin{eqnarray} \label{LandauPotential}
    \mathcal F_l(\theta, \eps) =  \frac{k_1(\theta -\theta_1)}{2} \eps^2  +
     \frac{k_2}{4} \eps^4  + \frac{k_3}{6} \eps^6,
    \end{eqnarray}
where $k_{1}$, $k_{2}$, and $k_{3}$  are  material-specific constants,  $\theta_1$
is the reference transformation temperature.

The non-local free energy function is usually constructed in a way similar to the class of
ferroelastic materials as follows \cite{Bubner1996,Bubner2000}:
    \begin{eqnarray}
     \mathcal F_g (\nabla \eps) = \frac{1}{2}k_g (\D{x}{\eps})^2,
    \end{eqnarray}
where $k_g$ is a material-specific constant. The non-local term above accounts for
inhomogeneous strain field. It represents energy contributions from domain walls of different phases (hence, an analogy with the Ginzburg term in semiconductors).
In order to account for
dissipation effects accompanying phase transformations, a Rayleigh dissipation term is
introduced here as follows \cite{Bales1991}:
    \begin{eqnarray}
    \mathcal F_R = - \frac{1}{2}\nu (\D{t}{\eps})^2,
    \end{eqnarray}
where $\nu$ is a material-specific constant. The above dissipation term accounts for the
internal friction. At macroscale, it is translated  into the viscous effects \cite{Abeyaratne2000}.

By substituting the potential energy function into the Lagrangian function given in
Equation~(\ref{Lagrange}),  and using the extended Hamilton's principle,  the governing
equation for the dynamics of the mechanical field can be obtained as follows:
    \begin{eqnarray}  \label{MechEq}
    \rho \ddot u = \D{x}{}\left ( k_1(\theta -\theta_1) \D{x}{u}  +
     k_2 \left(\D{x}{u}\right)^3  + k_3 \left(\D{x}{u}\right)^5 \right )
      + \nu \D{t}{}\DD{x}{u}  -  k_g \frac{\partial ^4u}{\partial x^4}.
    \end{eqnarray}

In order to formulate the governing equation for the thermal field, the conservation law of the
internal energy is employed here:
    \begin{eqnarray} \label{IntEng}
    \rho \D{t}{e} + \D{x}{q}  - \sigma \D{t}{\eps} - \nu \D{t}{\eps}\D{t}{\eps}
     - k_g \frac{\partial^2 \eps}{\partial x \partial t}  =0,
    \end{eqnarray}
where $e$ is the internal energy,  $q = -k\partial  \theta/ \partial x$ is the (Fourier) heat
flux, $k$ is the heat conductance coefficient of the material, and $\sigma$ is the stress.
In order to capture the coupling between the mechanical and thermal fields, the internal
energy is associated with the non-convex potential energy  mentioned above via the
Helmholtz free energy function as follows \cite{Falk1980,Matus2004,Bubner1996}:
\begin{eqnarray}
\mathcal H(\theta, \eps)= \mathcal F  - c_v \theta \ln \theta
\end{eqnarray}
and  the thermodynamical equilibrium condition gives:
    \begin{eqnarray}
         e = \mathcal H - \theta \D{\theta}{\mathcal H}, \quad \sigma = \D{\eps}{H},
    \end{eqnarray}
where $c_v$ is the specific heat capacitance. By substituting the above relationships into
Equation~(\ref{IntEng}), the governing equation for the thermal field can be formulated as:
    \begin{eqnarray} \label{ThermEq}
    c_{v}\D{t}{\theta}= k\DD{x}{\theta}+k_{1}\theta \eps \D{t}{\eps}
     + \nu \left ( \D{t}{\eps} \right)^2.
    \end{eqnarray}

It is shown clearly from the above derivation that the governing equations for both the
mechanical and thermal fields are constructed using the same potential function
 $\mathcal F_l(\theta, \eps)+ \mathcal F_g (\nabla \eps)$, which is taken here as the
 Ginzburg-Landau free energy function. It has been shown in
 Ref.(\cite{Bales1991,Melnik2002,Niezgodka1991})
 that the mathematical model such as the one based Equation~(\ref{MechEq}) and (\ref{ThermEq})
is capable  to capture the first order phase transformations in ferroelastic materials, and the
thermo-mechanical coupling, by suitably choosing the coefficients according experiment results
\cite{Bubner1996,Bubner2000,Niezgodka1991}, as shown in section 3.
The thermo-mechanical coupling is also shown by the term $k_1(\theta-\theta_1)\eps$, which
 accounts for the influence of the thermal field on the mechanical one,  and the other two terms
 $k_{1}\theta \eps \D{t}{\eps} + \nu \left ( \D{t}{\eps} \right)^2$ account for the conversion
 of mechanical energy into thermal form.
Due to strong nonlinearity in the mechanical field  and nonlinear coupling between
the mechanical and thermal fields, numerical simulations based on this model is far from trivial
\cite{Matus2004,Bubner1996}.

For the sake of convenience of numerical analysis discussed later on, the model is recast into
a differential-algebraic system in a way similar to  \cite{Matus2004,Melnik2002}:
\begin{equation} \label{SMAEq}
  \begin{array}{l}
 \displaystyle
 \rho\DD{t}{u} =  \D {x}{\sigma} -  k_g  \D{x^4}{^4u} +   F,  \\
 \displaystyle
c_{v}\frac{\partial\theta}{\partial
t}=k\frac{\partial^{2}\theta}{\partial x^{2}}+k_{1}\theta \eps
\D{t}{\eps}  + \nu \left ( \D{t}{\eps} \right)^2 + G, \\
 \displaystyle
\sigma^* =  k_{1}(\theta-\theta_{1}) \eps  -  k_{2}\eps^{3}+k_{3}\eps^{5} + \nu \D{t}{\eps}.
\end{array}
\end{equation}
where $u$ is the displacement, $\theta$ is the temperature, $\rho$ is the density,
$k_{1}$, $k_{2}$, $k_{3}$, $c_{v}, k_g, \nu$ and $k$ are  normalized
material-specific constants,  $\sigma^*$ is the effective stress incorporating the
viscous effects, and $F$ and $G$
are distributed mechanical and thermal loadings, which are all zero
for the current problem.

To couple the mathematical model for the  SMA rod and  that of the  block,
  boundary conditions for Equation~(\ref{SMAEq}) should be associated with
 the movement  of the mass block, which can be formulated as follows:
    \begin{equation}
    \begin{array}{l} \displaystyle
     \D{x}{\theta} = 0, \quad  u = 0, \quad  \textrm{at}~x=0, \\
  \displaystyle
  \D{x}{\theta} = 0,   \quad   \sigma(L) =  \frac{f_m}{\beta},
      \quad  \textrm{at}~x=L.
   \end{array}  \label{CoupleEq}
 \end{equation}

It is worthy to note that the above boundary conditions are not sufficient for
the system to have a unique solution. Indeed, when the nonlinear
constitutive law is used, as given in the last line of Equation~(\ref{SMAEq}), the
relationship between $\sigma$ and $\eps$ is no longer a one to one map.
For a given stress value,  one might be able to find three possible strain values
which all satisfy the constitutive law. Therefore, an additional boundary condition
 on  $u$ or $\eps$ is necessary to guarantee uniqueness. Here we employ the idea
given in Ref. \cite{Bubner1996} and set the additional boundary condition as
$ \partial \eps / \partial x = 0 $.  It is shown in Ref.(\cite{Niezgodka1991,Abeyaratne2000})
that, provided with such boundary conditions, the model has an unique solution and
can be numerically analyzed.

The geometrical coupling can be written as $x_m = L + u(L)$, so the velocity
of the mass block can be written as $v_m= \dot x_m  = \dot u(L)$. By putting
the two mathematical models, Equation~(\ref{MassEq}) and Equation~(\ref{SMAEq}),
and their coupling conditions  together,  our mathematical model
 can be formulated finally as:
\begin{equation} \label{OverallEq}
  \begin{array}{l}
     \displaystyle
 \rho\DD{t}{u} =  \D {x}{\sigma}-  k_g  \D{x^4}{^4u},  \\
     \displaystyle
c_{v}\frac{\partial\theta}{\partial
t}=k\frac{\partial^{2}\theta}{\partial x^{2}}+k_{1}\theta \eps
\D{t}{\eps} +  \nu \left ( \D{t}{\eps} \right)^2, \\
     \displaystyle
\sigma^* =  k_{1}(\theta-\theta_{1}) \eps  -  k_{2}\eps^{3}+k_{3}\eps^{5} + \nu \D{t}{\eps}, \\
     \displaystyle
 \D{x}{\theta} =  0,  \quad  u = 0, \quad  \D{x}{\eps}=0, \quad  \textrm{at}~x=0, \\
      \displaystyle
 \D{x}{\theta} =  0,  \quad   \sigma = \frac{1}{\beta} ( m \ddot u(L)
    + \mu_m \dot u(L) + k_m u(L) ), \quad
   \D{x}{\eps}  = 0,    \quad   \textrm{at}~x=L.
\end{array}
\end{equation}

The above model incorporates the phase transformation, thermo-mechanical coupling,
 viscous effects, and interfacial contributions, it is natural to expect that the model is
 able to model the vibration damping effects due to these contributions.


\section{Vibration Damping Due to Phase Transformation}

As mentioned earlier,  damping effects in SMA rod can be attributed to several factors.
The internal friction (viscous effect) contributes to the damping effects when phase
transformation is taking place.
 Since the occurence of  thermo-mechanical coupling, the SMA materials also have damping effects
 due to the energy conversion between thermal and mechanical fields.
The most pronouncing contribution to the damping effects in the SMA rod is the mechanical
hysteresis due to phase transformation between martensite variants. Unlike other contributions
to the damping effects, the contribution due to hysteresis can be graphically explained using
the non-convex potential energy function and the non-convex constitutive relation.

\begin{figure}   \begin{center}
  \includegraphics[scale=0.4 ]{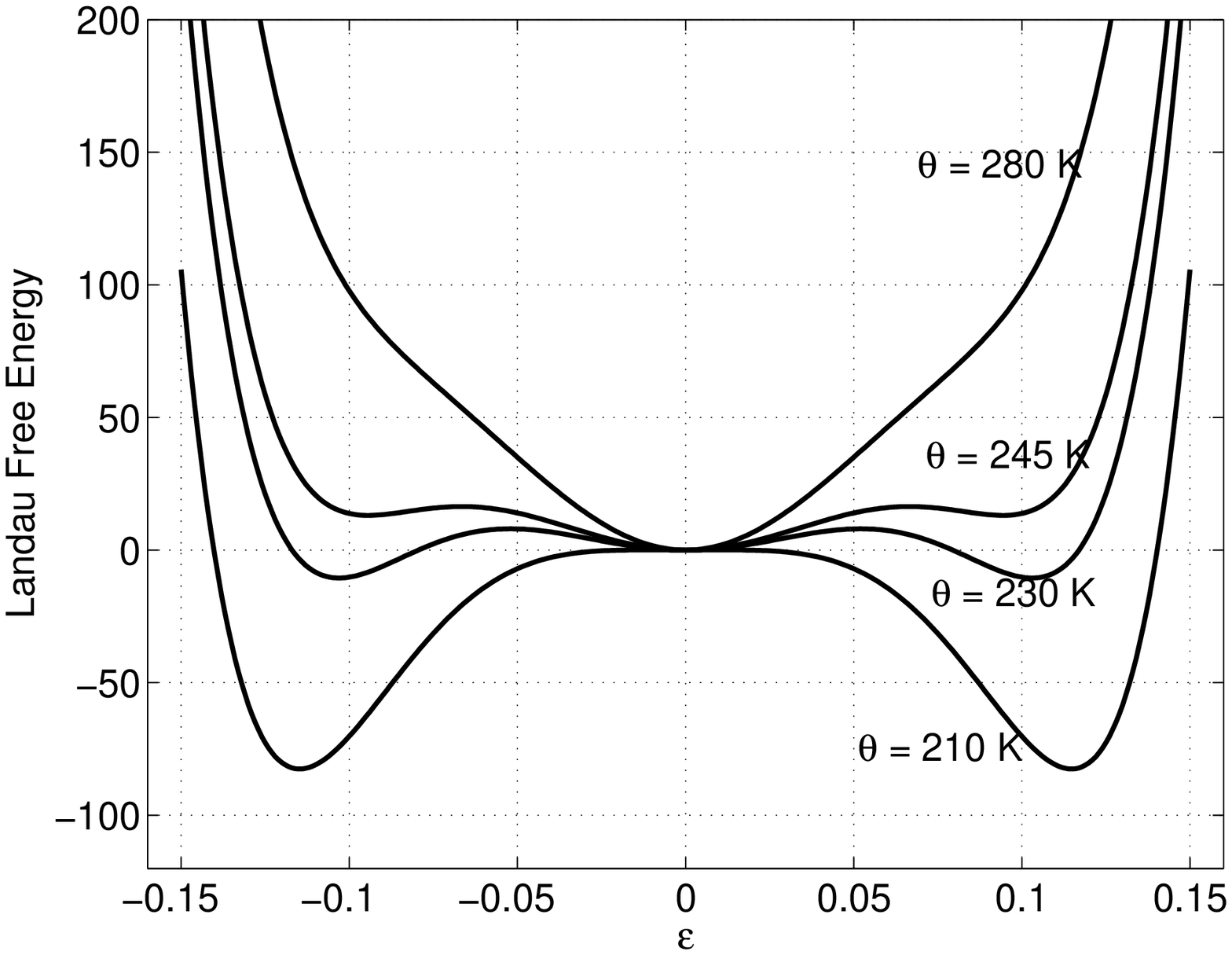}
  \includegraphics[scale=0.4 ]{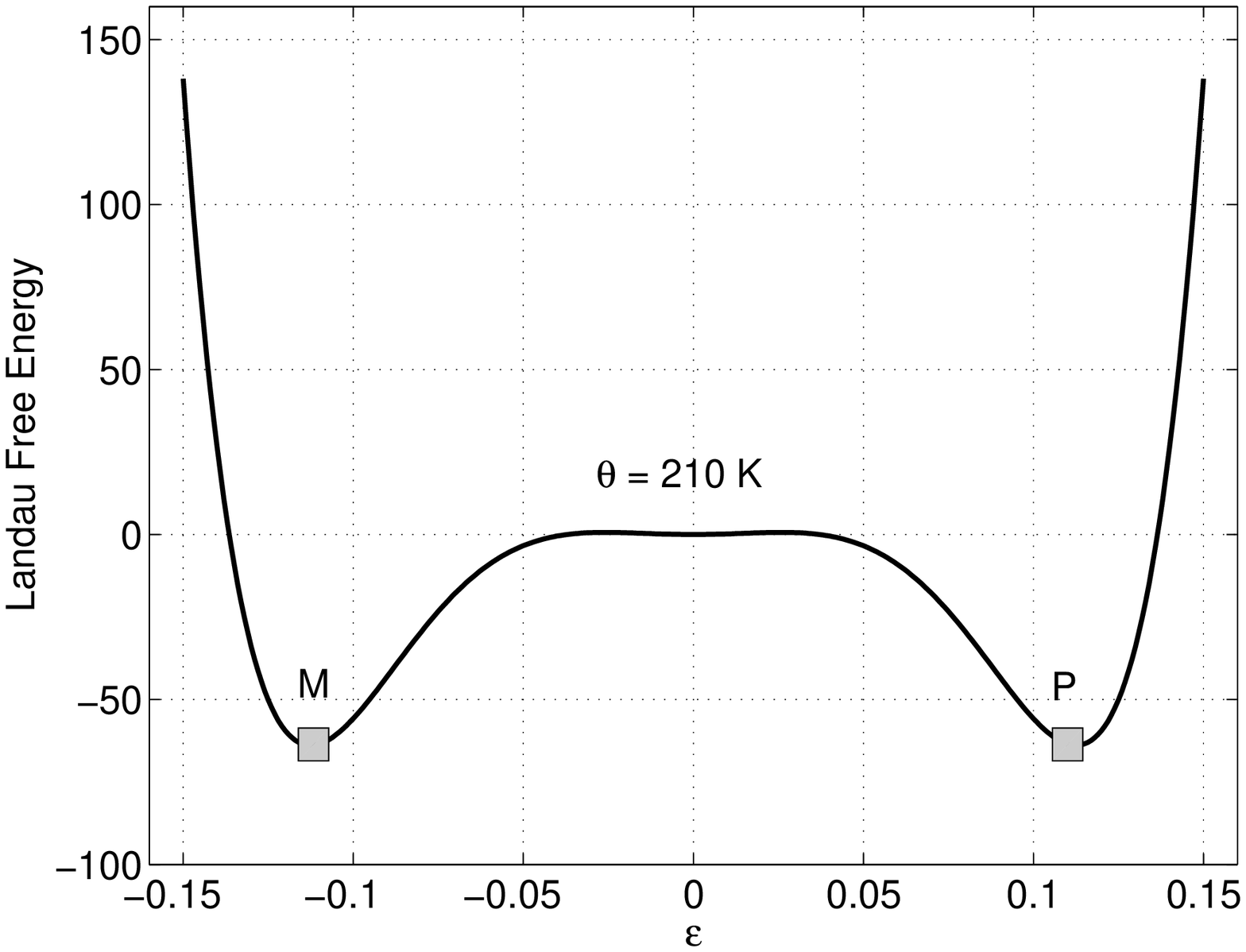}
 \caption{Non-convex free energy and its dependence on temperature}
 \label{LandauEnergy}
 \end{center}    \end{figure}

In the heart of the Ginzburg-Landau theory for the first order phase transformation in SMA, lies
the free energy function defined as a non-convex
function of the chosen order parameter for characterization of all the phases and variants
(e.g., \cite{Bubner1996,Falk1980,Kloucek2004,Wang2004}), which is defined as Equation
(\ref{LandauPotential}).   One example of the Landau free energy function is shown in Figure
 (\ref{LandauEnergy}), for which the material and its physical parameters are given in section 5.
It is shown clearly in the figure that the function has only one local minimum at high
 temperature ($\theta=280$K), which is associated with the only stable phase (austenite)
 in this case. At lower temperature ($\theta=210$K), there are two symmetrical local minima
 (marked with small gray rectangles), which are associated with martensite plus (P)
and minus (M), respectively.  For vibration damping
purpose, it is always beneficial to make the SMA work at low temperature, so the hysteresis
loop will be larger.

\begin{figure}   \begin{center}
 \includegraphics[scale=0.4] {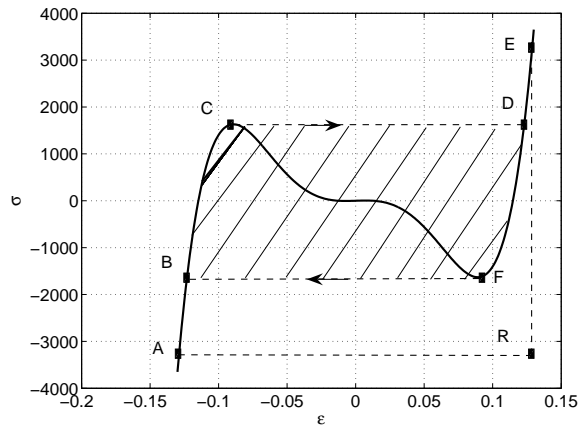}
\caption{hysteretic behaviour and its damping effect caused by the non-convex free energy}
 \label{Hysteresis}
 \end{center} \end{figure}

The hysteretic behaviour and mechanical energy dissipation of the SMA rod at low temperature
can be explained using its nonlinear constitutive relation as shown in  Figure  (\ref{Hysteresis}),
which is obtained using the thermodynamical equilibrium condition \cite{Bubner1996}:
	\begin{eqnarray}
         \sigma = \D{\eps}{\mathcal F}
	\end{eqnarray}

For a given external loading, it will induce an internal stress and deformation in the SMA rod,
part of the work done by  the loading is stored in the rod via its deformation
and  can be calculated as:
    \begin{equation}        \label{Work}
       W =  \int_{\eps_s}^{\eps_e} (- \sigma \eps) d \eps,
    \end{equation}
\noindent where $\eps_s$ and $ \eps_e$ are starting and ending strain values, respectively.

For pure elastic material, the stored energy can be released without loss when the external
loading is removed, there is no mechanical energy dissipation. For  thermo-mechanical coupling material,
the stored energy is also obtainable when the loading is removed, but part of the input energy
will be converted into thermal form and can not be fully released. If the material is insulated, its
temperature will be slightly increased due to the input of mechanical energy. When
viscous effects of the material are taken into account,
the fraction of mechanical energy converted into thermal form will be larger, which means that
dissipation of mechanical energy into thermal form is enhanced by viscous effects.
But, compared to those due to mechanical hysteresis, all these energy dissipation are
not remarkable.

The energy dissipation due to hysteresis can be presented schematically by considering
 the mechanical energy loss (converted from mechanical to thermal) in  one cycle of loading
 \cite{Masuda2002,Kloucek2004}.  Assume that
 the loading process starts from point $A$ in Figure~(\ref{Hysteresis}), and  the load is
  continuously increased to point $E$. Then,
the constitutive relation in this case will be the curve $ABCDE$, and  the
work done by the loading is represented by the area of $ABCDERA$. When the load
is decreased from $E$ to $A$,  the system will follow another
constitutive curve $EDFBA$,  and the work done by the rod is represented by  the area
of $EDFBARE$.  There is a difference between the two areas, which is enclosed by the
 curve $BCDFB$ and  hatched in the figure.  This area represents the mechanical
 energy loss due to the hysteretic behaviour. At the same time,  due to the thermo-mechanical
 coupling, the temperature of the material will be increased, as indicated by the source
 terms $k_1\theta \eps \D{t}{t}+ \nu \left ( \D{t}{\eps} \right)^2$ in the energy equation.

The physics behind this is that, when the external energy is applied to the SMA rod,
 part of the energy will be demanded by the material to convert
itself from one variant  (P) to another (M), as sketched in Figure (\ref{LandauEnergy}).
Both P and M are stable states,  and have the same potential,  so the material will not recover
its previous configuration when the loading is removed. In this sense, the SMAs have
plastic-like property.  Part of the
input mechanical energy is consumed just for  converting  the material from one variant to
 another (martensite phase transformation), and  finally converted into thermal energy via
  the thermo-mechanical coupling.
If one want to reverse the transformation, extra input energy is demanded.


\section{Numerical Methodology}

The constructed model for the vibration damping system given in
Equation~(\ref{OverallEq}) is a strongly nonlinear coupled system.
 Experiences gained from the numerical experiments on similar problems
 shows that the development of an efficient algorithm  is necessary for the
numerical analysis of the model because standard iteration algorithms
have difficulties to cope with the very strong nonlinearity
\cite{Bubner1996,Kloucek2004,Melnik2002,Wang2004}. For the current problem,
an additional difficulty is that the boundary conditions are system dependent.
Indeed, we observe that the stress on the right end of
the rod is a function of its displacement, velocity, and acceleration.
In order to deal with these difficulties,  the model is reduced to
 a differential-algebraic system and solved by using backward differentiation
 in time.
The practical implementation is done similarly to
\cite{Matus2004,Melnik2002,Wang2004}. Note, however, that here for the Chebyshev
pseudo-spectral approximation, a set of
Chebyshev points $\{x_i\}$ are chosen along the length direction as follows:
\begin{equation}
  \label{num-eq4}
  x_i  = L\left(1-\cos(\frac{\pi i}{N})\right) /2, \quad i=0,1,\dots,N.
\end{equation}

Using these nodes, $\eps, v, \theta$, and  $\sigma$ distributions in the rod can
be expressed in terms of the following linear approximation:
\begin{equation}
  \label{num-eq2}   f(x) = \sum_{i=0}^{N} f_i \phi_i(x),
\end{equation}

\noindent where $f(x)$ stands for $\eps, v, \theta$, or  $\sigma$,
 and $f_i$ is the function value at $x_i$. Function  $\phi_i(x)$ is the $i^{th}$ interpolating
polynomial which has the following property:
   \begin{equation}
     \phi_i(x_j) = \left \{
     \begin{array}{ll}
       1,  & i=j, \\  0, & i\neq j.
     \end{array}  \right.
   \end{equation}

It is easy to see that the Lagrange interpolants would satisfy the
 the above requirements.  Having obtained $f(x)$ approximately, the
derivative $\partial f(x)/ \partial x$ can be  obtained by
taking the derivative of the basis functions $\phi_i(x)$ with respect to $x$:
 \begin{equation}
   \label{num-eq3}
  \frac{\partial f}{\partial x} = \sum_{i=1}^{N} f_i
  \frac {\partial \phi_i(x)}{\partial x}.
\end{equation}

All these approximations are formulated in matrix form, for the convenience
of actual programming implementation. For approximation to higher order derivatives,
similar matrix form can be easily obtained.  By substituting all the approximation into
the DAE system, it will be recast into a set of nonlinear algebraic equations, which
can be solved by Newton-Raphson iteration, following a similar way as done in
Ref.~(\cite{Matus2004,Bubner1996,Melnik2002,Wang2004}).


\section{Numerical Experiments}

Several different numerical experiments have been carried out to demonstrate
the damping effects  of hysteresis in the SMA rod induced by the phase transformations.
All experiments  reported in this section have been performed on a
  $\textrm{Au}_{23}\textrm{Cu}_{30}\textrm{Zn}_{47}$ rod,
with length of 1$~cm$. Most of the physical parameters  for this specific material
 can be found in the literature (e.g., \cite{Falk1980,Melnik2002,Niezgodka1991,Wang2004}),
 which are listed as follows for the sake of convenience:
\begin{eqnarray*}
 k_{1}=480\, g/ms^{2}cmK, \quad  k_{2}=6\times10^{6}g/ms^{2}cmK,\qquad
  k_{3}=4.5\times10^{8}g/ms^{2}cmK,   \\
\theta_{1}=208K,\quad  \rho=11.1g/cm{}^{3},\quad
c_{v}=3.1274g/ms^{2}cmK, \quad k=1.9\times10^{-2}cmg/ms^{3}K.
\end{eqnarray*}

There are two coefficients, $k_g$ and $\nu$, are not easy to find its value.
Normally, $k_g$ is very small and here we take it as $5$ by referring to the value indicated in
Ref.(\cite{Bubner1996,Bubner2000}). The value of $\nu$ is take as $10$, with
the same unit system. To demonstrate the damping effects of the SMA rod,
we set $\mu_m=k_m=0$ in Equation~(\ref{OverallEq}). Then, only the force from
the SMA rod is taken into account.

For all numerical experiments, the initial conditions for the rod are chosen  as
 $\theta = 210^o$, $\eps=0.115$,  $u=v=0$,   the simulation  time span is $[0,4] ms$,
and $40$ nodes  have been used for the Chebyshev approximation.

\begin{figure}
 \includegraphics[scale=0.4]{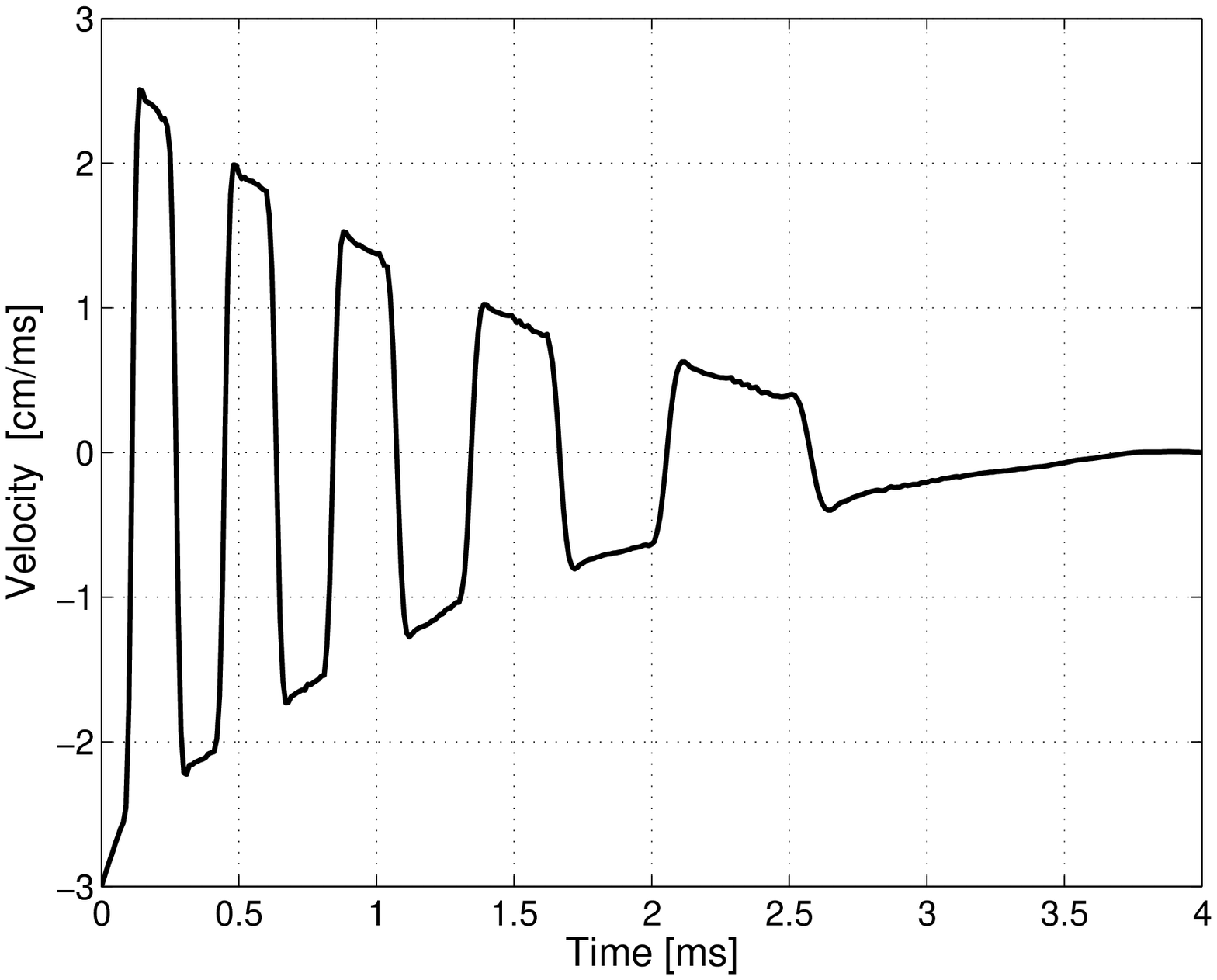}
 \includegraphics[scale=0.4]{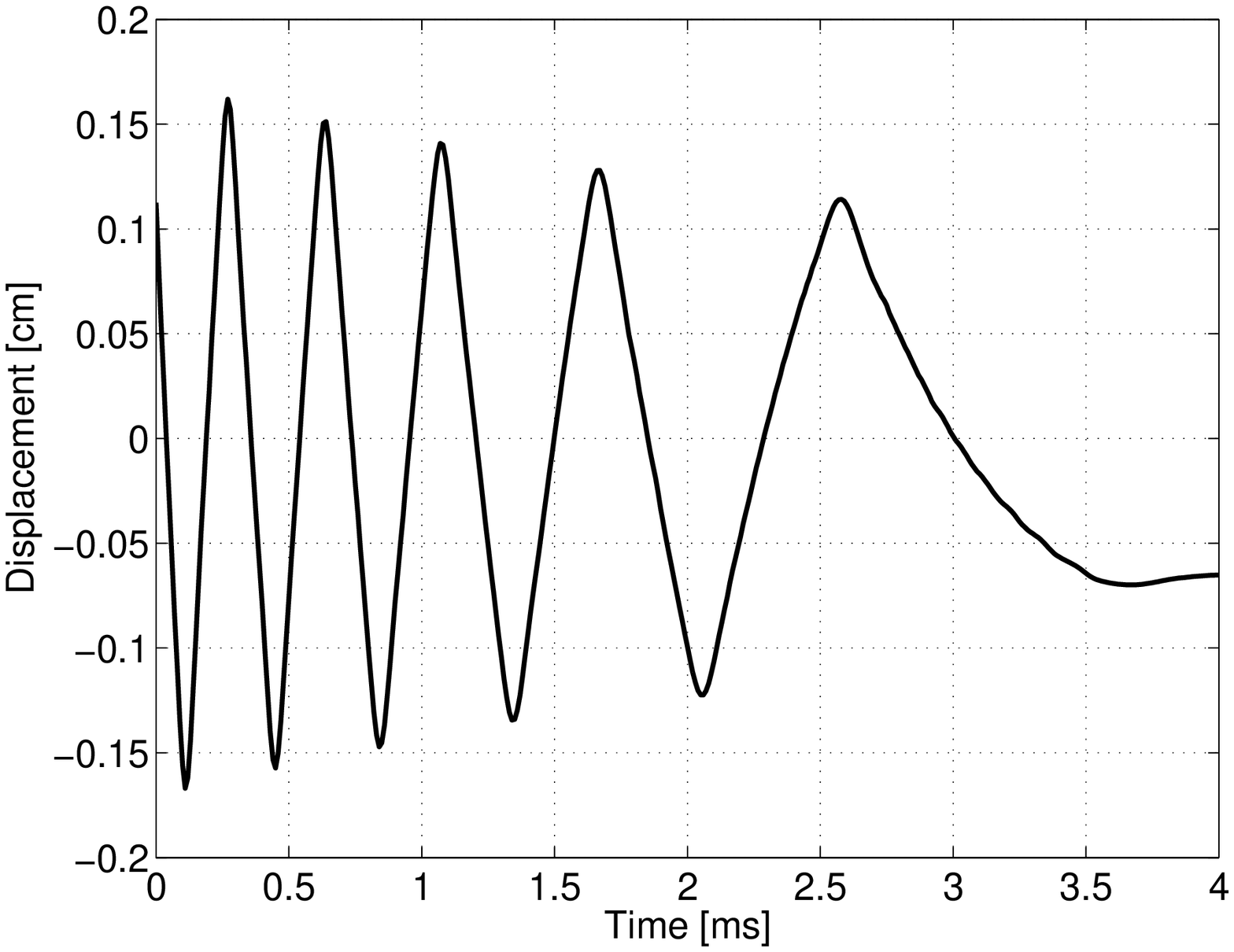}   \\
 \includegraphics[scale=0.4]{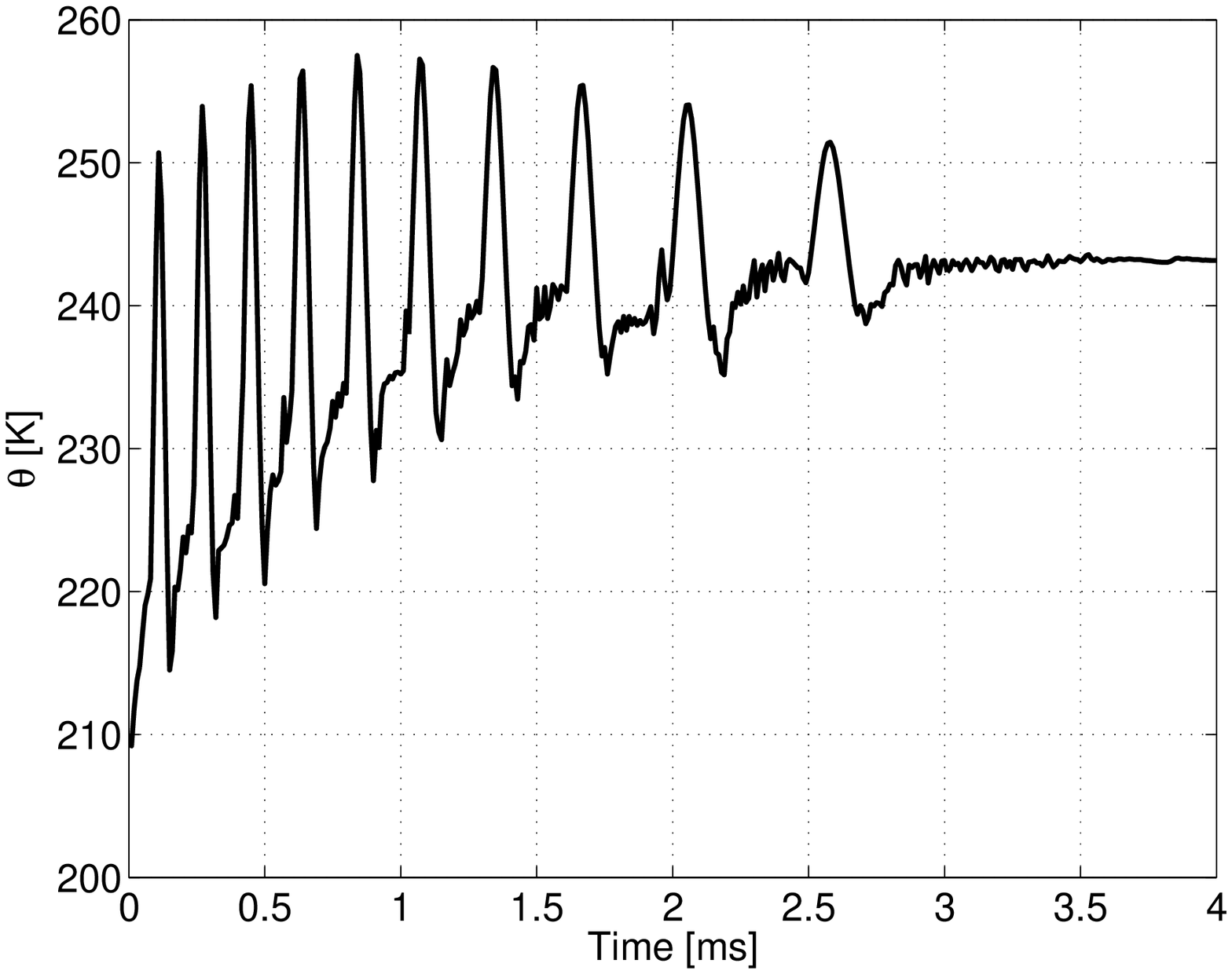}
 \includegraphics[scale=0.4]{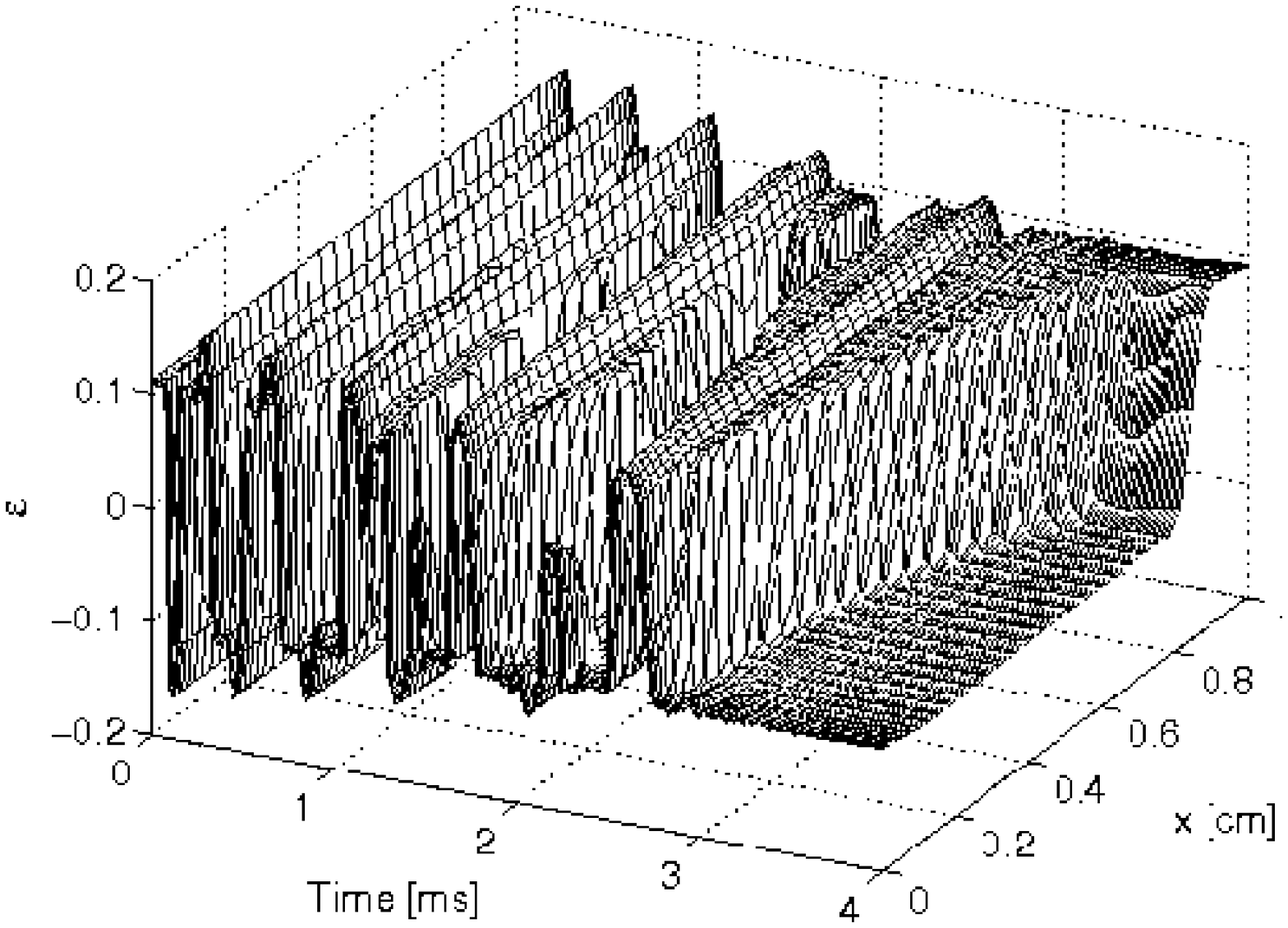}
\caption{Damping effect of a SMA rod involving mechanically induced phase
transformations}
 \label{NumExp1}
  \end{figure}

In the first experiment reported here, the parameter $m/\beta$ is chosen as $200~(g/cm^2)$
and the initial value of $v_m$ is chosen $-3~(cm/ms)$.  The numerical  results are
presented in Figure~(\ref{NumExp1}).

From the upper left plot it can be observed that the velocity of the block is damped effectively,
and gradually the velocity tends to a small value.  The entire damping process can be roughly
 divided into two stages. The first stage is $t\in [0,2.6] ms$ in which the damping is more
 effective since the vibration energy is still large enough to switching the whole SMA rod
between  martensite plus and minus at low temperature (martensite phase transformation).
While in the the second stage,
 the contribution from phase transformation becomes weaker and finally fades because
 the temperature is raised, the mechanical energy is also dissipated and not large enough to
 induce phase transformation. The motion of the rod finally becomes a
thermo-mechanical vibration without phase transformations at the end of the simulation.
The strain distribution, presented in the  lower right  part of the plot, shows this clearly.
In the lower left plot, we present the average temperature of the rod, which could be
regarded as a measurement of thermal energy. Since insulation thermal conditions are used
 and there is no thermal dissipation included in the model, the temperature is increased rapidly
 when there is phase transformation induced because the large change rate of strain, and much
 slower when there is no transformation.
This observation indicates that  the conversion of mechanical energy into thermal form
 is much faster when phase transformation  takes place. The oscillation of average temperature
 is caused by the phase transformation and thermo-mechanical coupling.

\begin{figure}
  \includegraphics[scale=0.4]{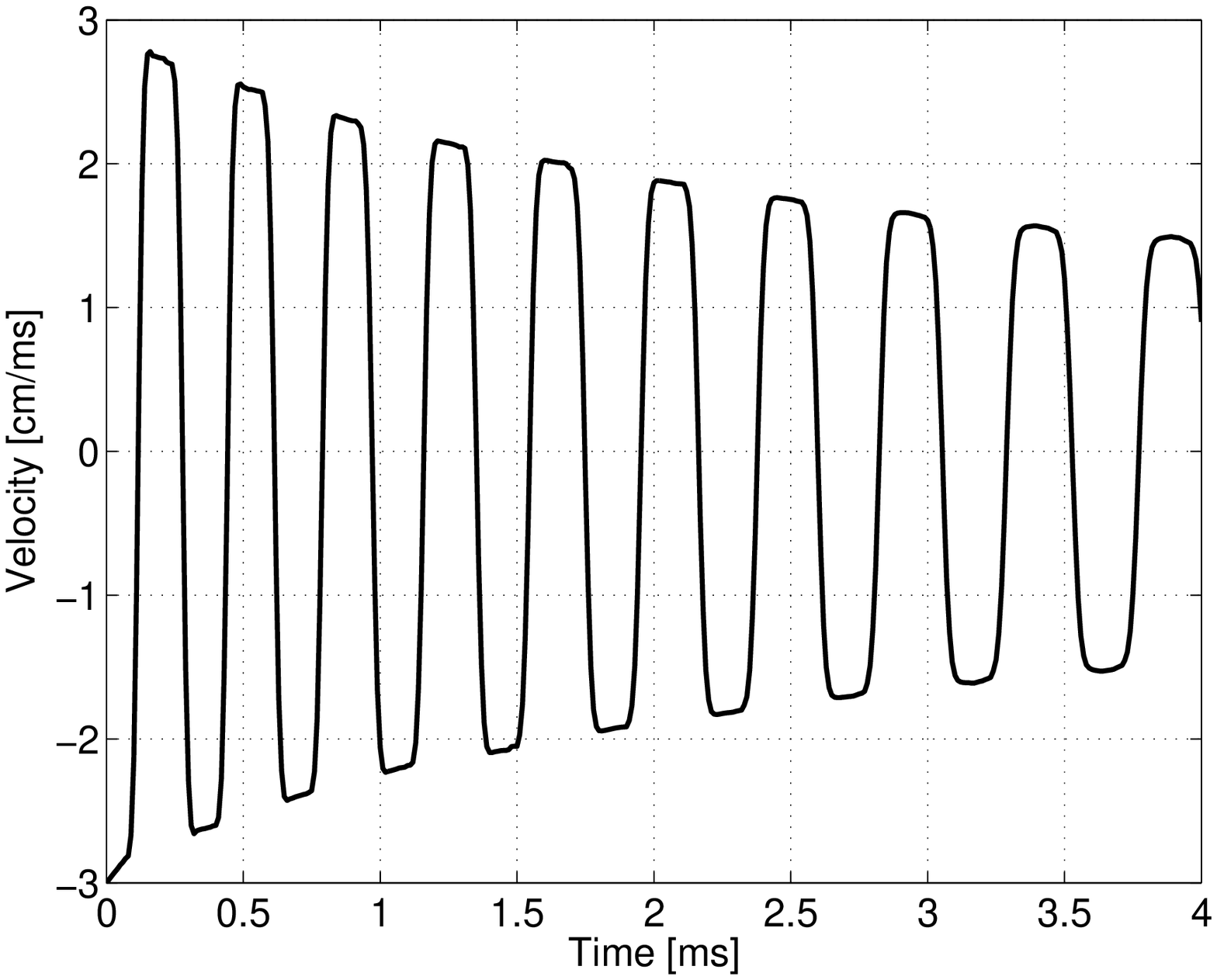}
  \includegraphics[scale=0.4]{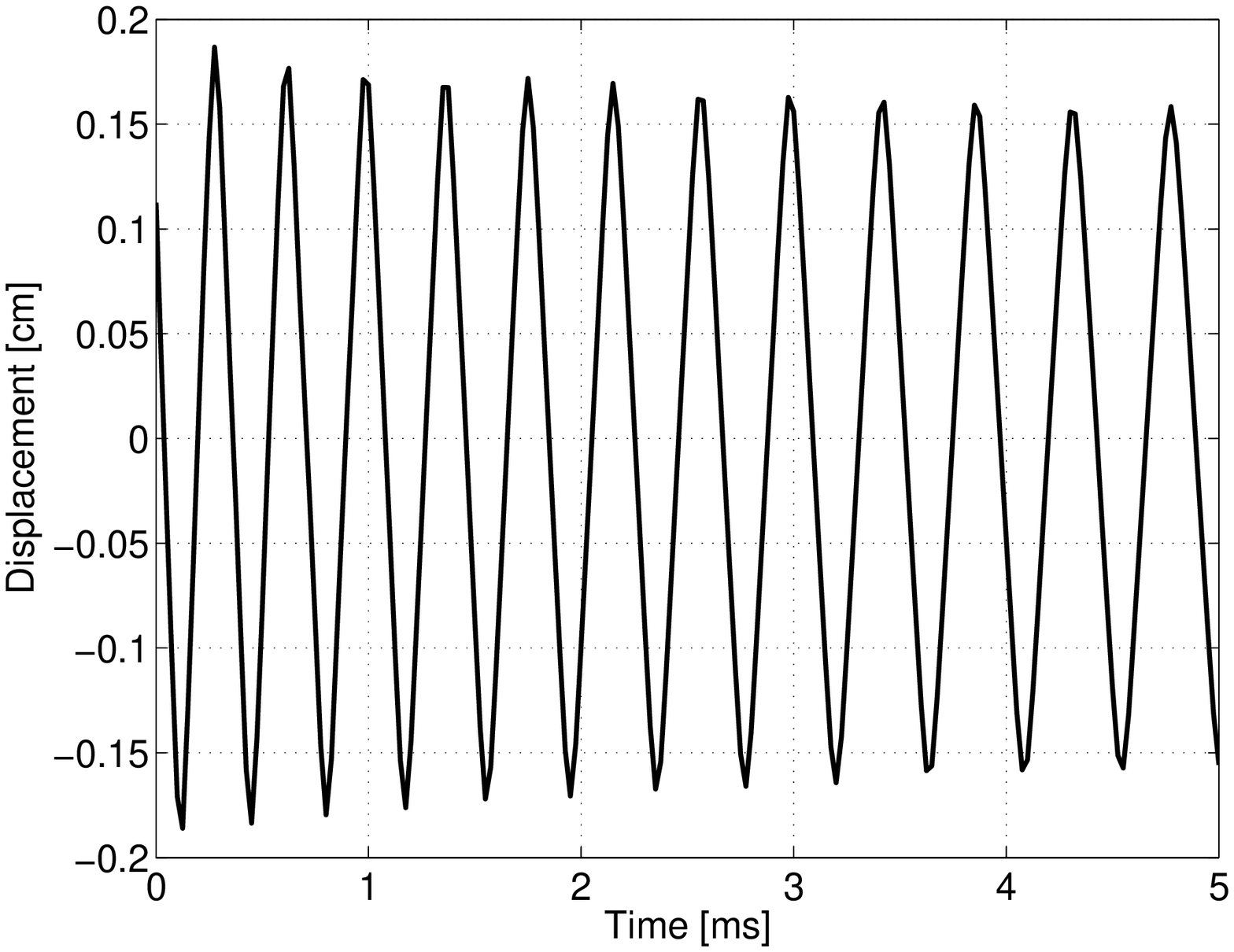} \\
  \includegraphics[scale=0.4]{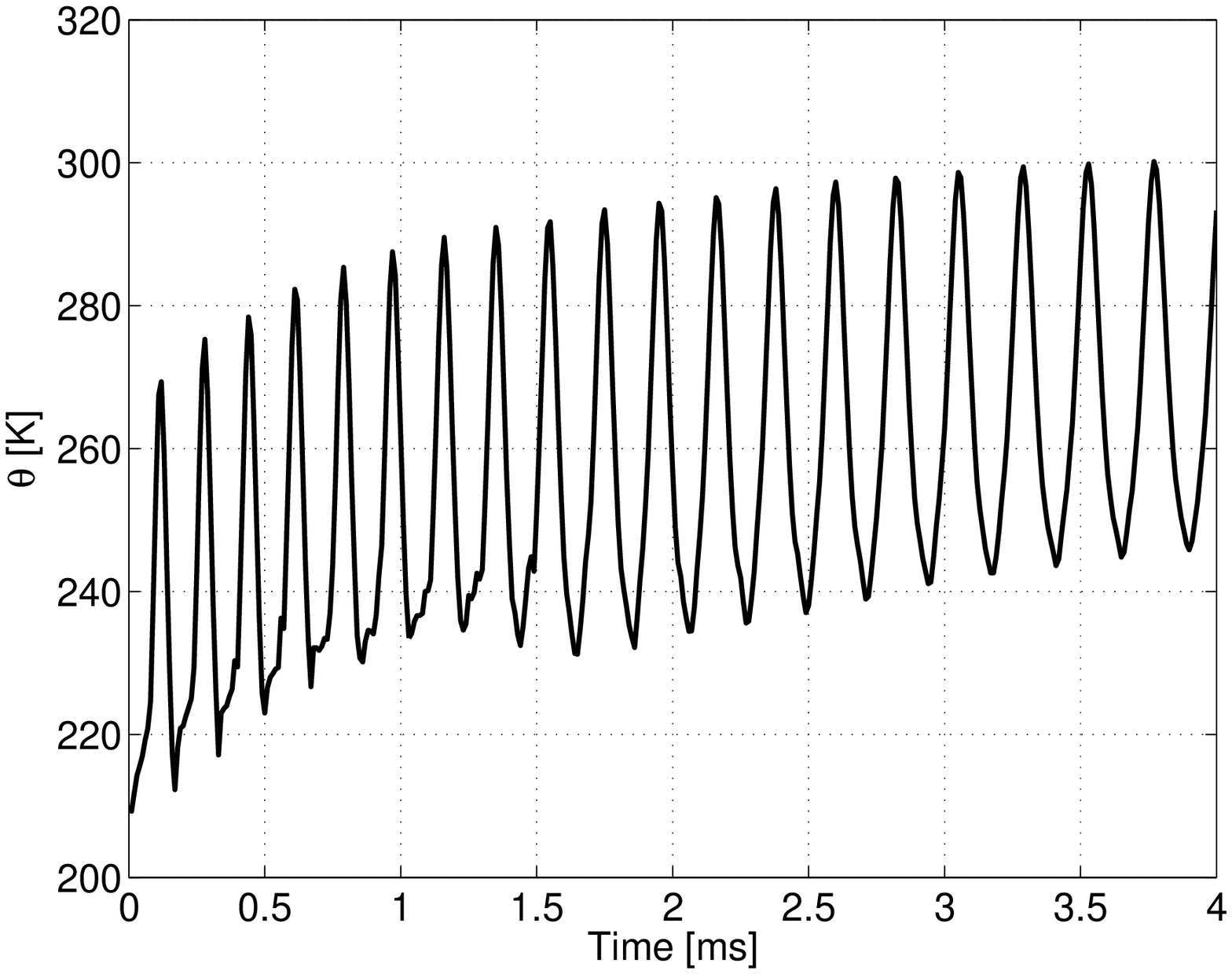}
  \includegraphics[scale=0.4]{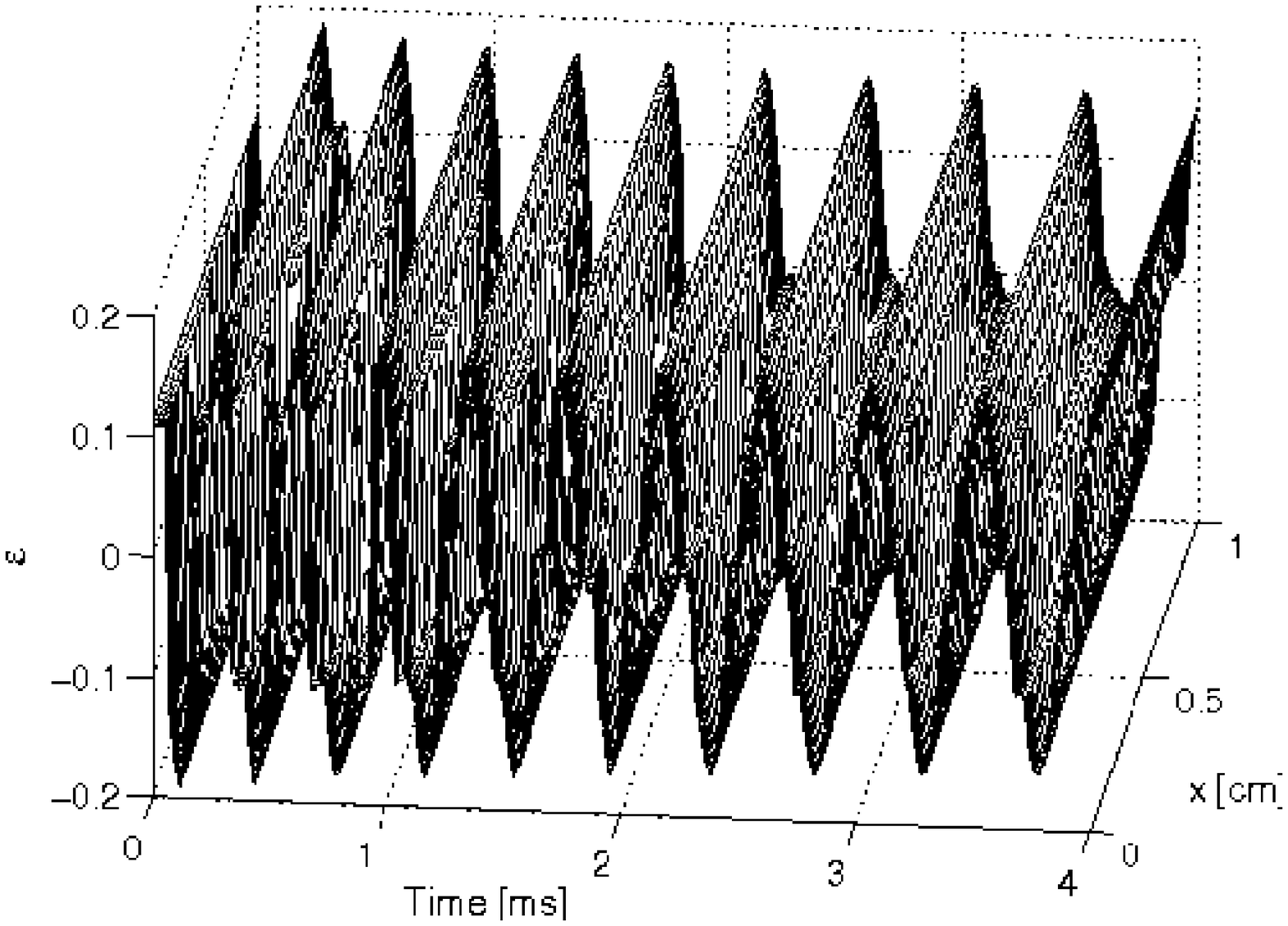}
 \caption{Damping effect of a SMA rod with large initial vibration energy}
  \label{NumExp2}
   \end{figure}

To analyze the  damping effects further, we have increased the initial vibration energy
 to $m / \beta = 500$.  The numerical results for this case are  presented in
 Figure~(\ref{NumExp2}), in a similar way as those in  Figure~(\ref{NumExp1}).
 It is seen from the strain plot that the whole rod is switched between compressed ($\eps <0$)
 and stretched ($\eps>0$) states in the entire simulation range, and $v_m$ is
 decreased continuously. This strain switching causes the temperature oscillation.
 The dissipation of mechanical energy is faster at the beginning and becomes slower after
 a while in the simulation,
 which is indicated by the velocity plot (consider that mechanical energy is proportional
 to the square of velocity). This is due  to the fact that phase transformation only takes
 place in  a short period in which the rod  temperature is still low, which can be roughly
 read as $[0,1.4]ms$ form the temperature plot, in which the temperature increases faster.
Afterwards,  there is no phase transformation since the rod is at high temperature state and
 the  motion becomes thermo-mechanical vibration again. This observation leads to same
 conclusion that phase  transformation  contribute most to the damping effects
 (energy conversion).

\begin{figure} \begin{center}
 \includegraphics[scale=0.4]{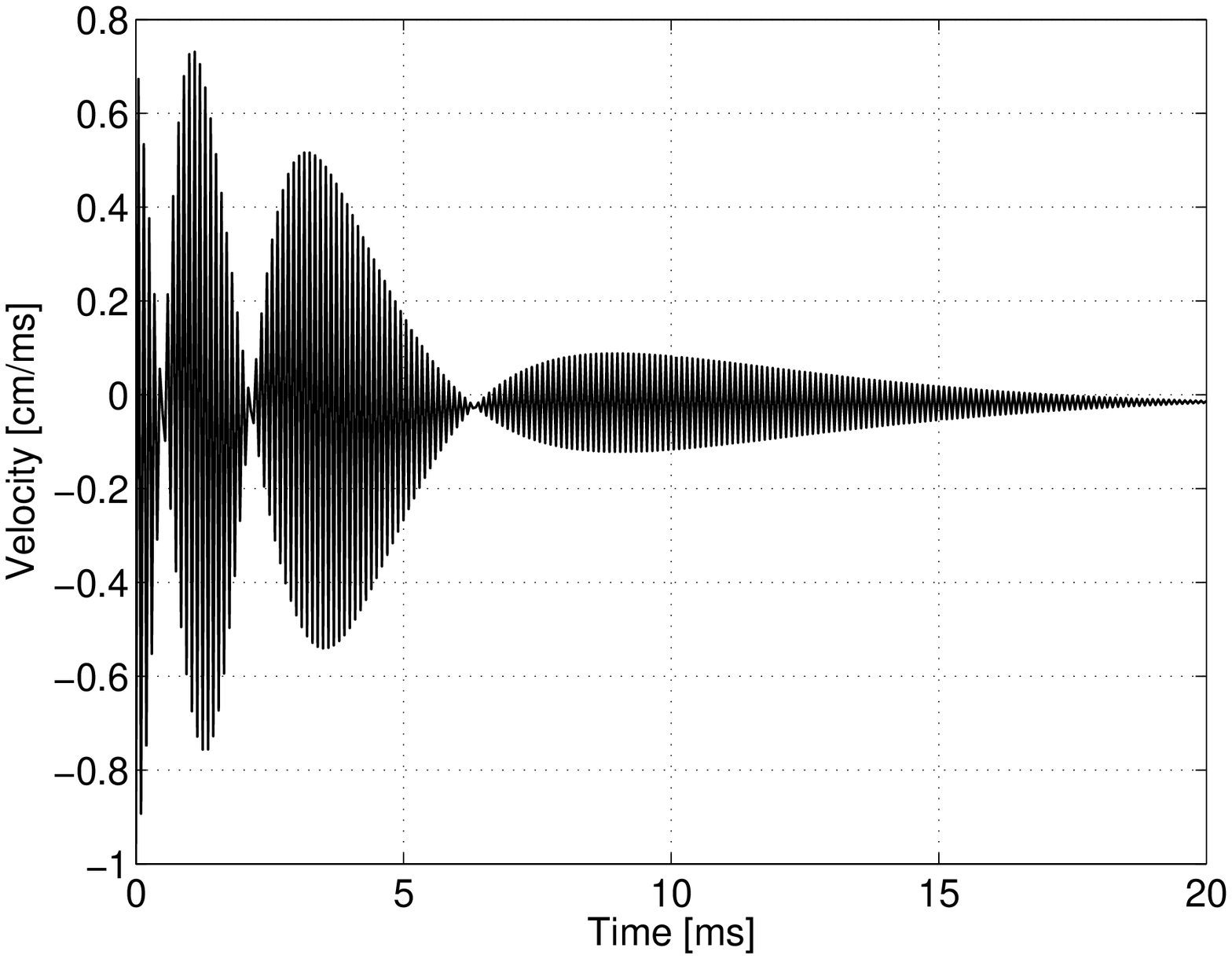}
 \includegraphics[scale=0.4]{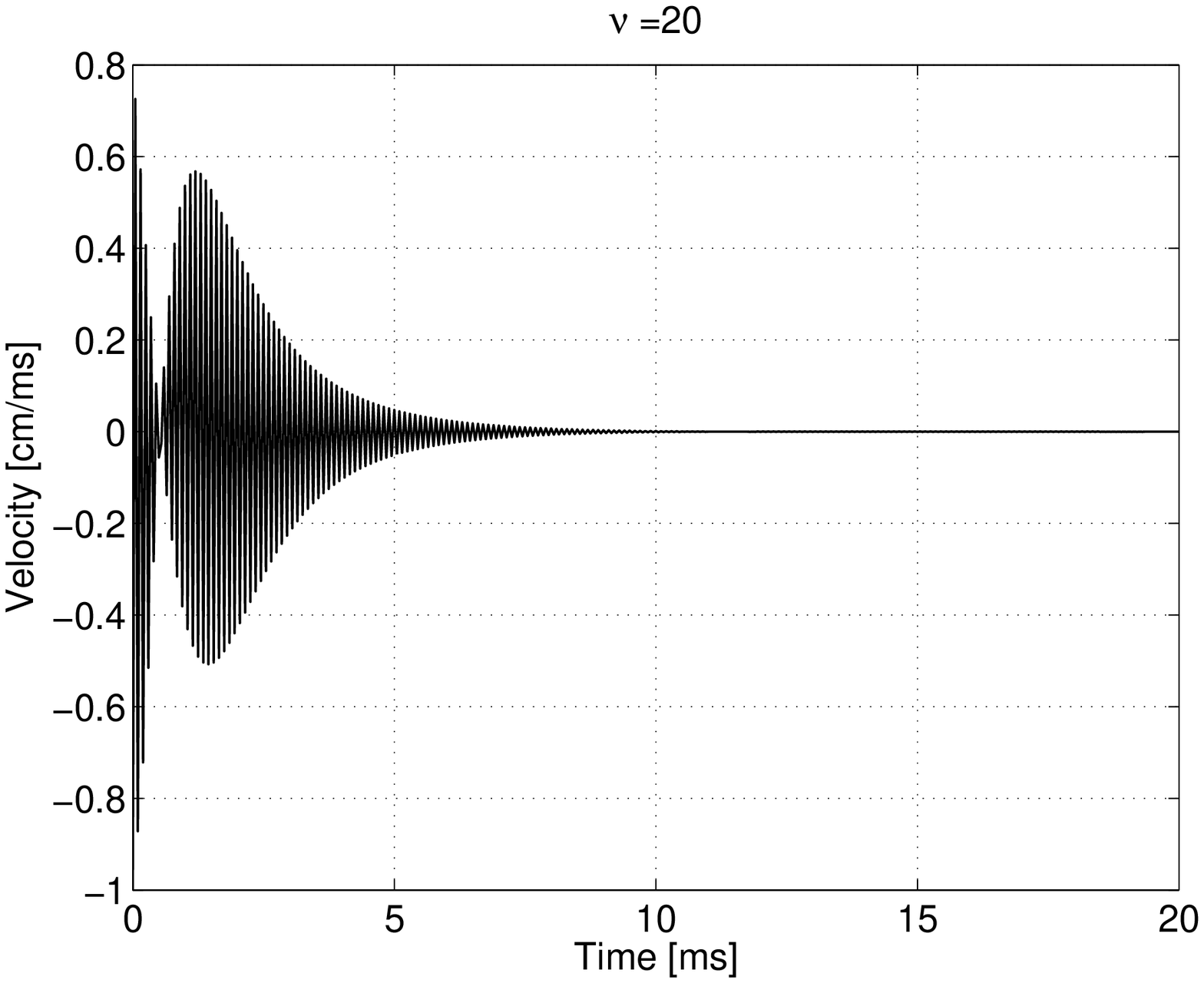}   \\
 \includegraphics[scale=0.4]{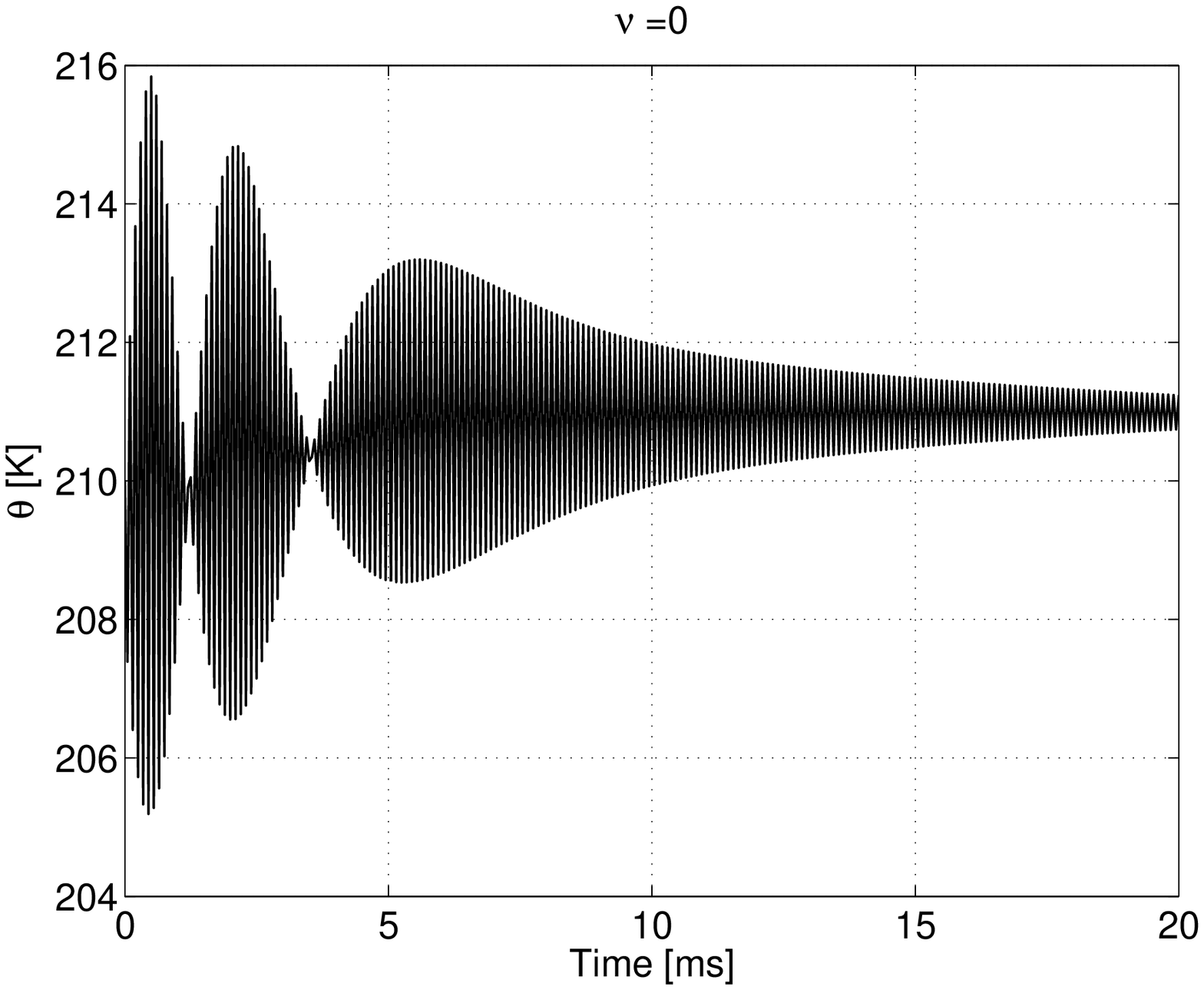}
 \includegraphics[scale=0.4]{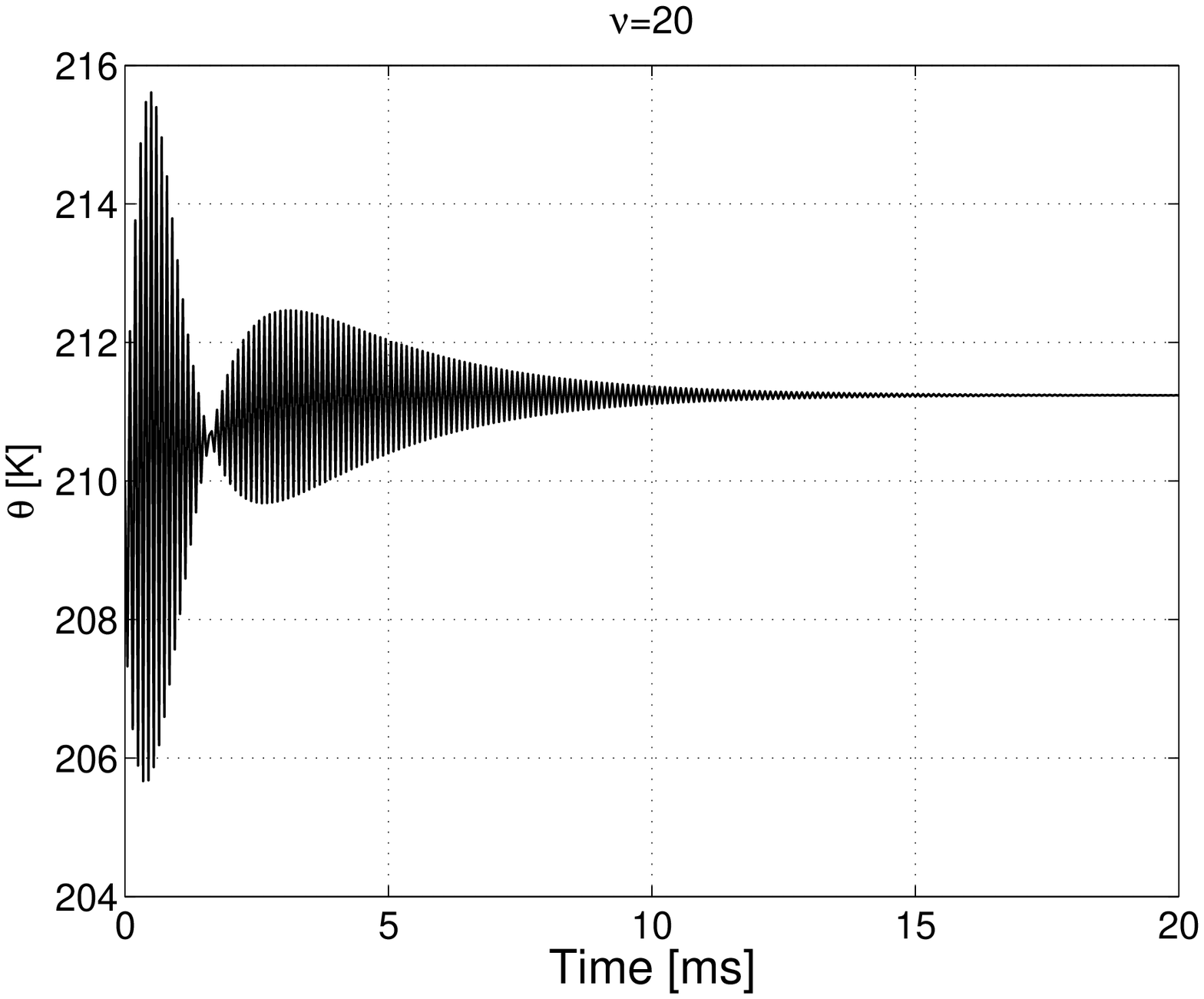}
  \caption{Effects of internal friction on damping performance of shape memory
  alloy rods (no phase transformations)}
  \label{DisspEffect}
  \end{center}  \end{figure}

To show the damping effects contributed from the viscous effects and thermo-mechanical
coupling, the last numerical example presented
here has been dealing with the analysis of the damping effects when there is no phase
transformation induced in the SMA rod at all.  For this purpose, the parameters are chosen as
 $m/\beta= 20$,  and initial velocity $v_m=-1$ such that the input mechanical energy is too
 small to induce phase transformation, and the simulation time span is also extented to
 $[0,20]$ms to capture the long time behaviour since the mechanical energy dissiption might be
 much slower.   For the simulation of damping effects contributed from thermo-mechanical
 coupling alone, the parameter is set  $\nu = 0$ (no viscous effects), and the numerical results
are presented on the left column of  Figure~(\ref{DisspEffect}), by plotting the evolution of
mass block velocty (top) and average temperature of the SMA rod (bottom).
It indicates that the coupling alone does contribute to damping effects, the mechanical energy
stored via the mass block motion is finally  converted into thermal form, and increase the
SMA rod temperature slightly. But, the dissipation of mechanical energy into thermal form is
rather slow, there are still oscillations at the end of the simulation.
 In order to include the contribution from viscos effects, now the viscosity parameter is set
 $\nu =20$, numerical results are presented on the right column similarly in the figure.
 It is shown clearly that  when viscous effects are incorportate  (right),
 the conversion of energy due to thermo-mechanical coupling is enhanced, the
the system is already at rest at $t=15$ms.  In both case, the oscillation of velocity and average
temperature has a series of peaks, which indicates a complicate nature of the
thermo-mechanical energy conversion.

In all these three numerical experiments, the energy conversion between mechanical and
 thermal form due to thermo-mechanical coupling is well captured.
It is shown that when the phase transformation is induced, the vibration damping
is remarkabley enhanced because the energy conversion is enhanced.  Because there is no
heat loss considered here, the average temperature of the rod  increases continuously.
It is clear that if the SMA damper is controlled at low temperature by external efforts, it will
be an effective damper.


\section{Conclusions}

In this paper, we constructed a mathematical model for vibration  damping
of a mass block connected with  a SMA rod.
We employed the modified Ginzburg - Landau theory for the mathematical modelling
 of the SMA rod to capture the phase transformation and thermo-mechanical coupling.
The model for the vibration of the block is coupled with the dynamics of the SMA rod
by adjusting the boundary conditions.  The model is then numerically analyzed and the
 damping characteristics of the SMA rod due to the first order martensite phase
 transformation  are investigated. It is shown that the vibration can be effectively damped,
 if the  first order phase transformation is induced.



\newpage

\newpage

\newpage

\newpage

\end{document}